\documentclass[sigconf]{acmart}

\AtBeginDocument{%
  \providecommand\BibTeX{{%
    \normalfont B\kern-0.5em{\scshape i\kern-0.25em b}\kern-0.8em\TeX}}}

\usepackage{caption}
\usepackage{subcaption}
\usepackage{array}
\usepackage{multirow}
\usepackage{booktabs}
\usepackage{soul}
\usepackage{graphicx}
\usepackage{environ}
\usepackage{xcolor}
\usepackage{colortbl}
\usepackage{framed}
\usepackage{verbatim}

\hypersetup{
    colorlinks=true,
    linkcolor=blue,
    filecolor=magenta,      
    urlcolor=cyan,
    citecolor=blue,
    pdfpagemode=FullScreen,
    }
 
\NewEnviron{makebold}{\textbf{\BODY}}
\NewEnviron{removemore}{\sout{\BODY}}
\colorlet{leftbarcolor}{gray!40}

\usepackage{enumerate}  
\usepackage[shortlabels]{enumitem}
\usepackage{makecell}
\usepackage{booktabs}
\usepackage{tabularx}
\usepackage{array}

\usepackage{xcolor}
\usepackage{soul}
\usepackage{ifthen}
\usepackage{xcolor}
\usepackage{listings}

\lstdefinestyle{promptstyle}{
  basicstyle=\ttfamily\small,
  columns=fullflexible,
  keepspaces=true,
  breaklines=true,
  breakatwhitespace=false,
  frame=single,
  framerule=0.4pt,
  rulecolor=\color{black!25},
  backgroundcolor=\color{black!2},
  numbers=left,
  numberstyle=\tiny\color{black!55},
  stepnumber=1,
  numbersep=10pt,
  tabsize=2,
  showstringspaces=false
}

\newboolean{useColorMapping}
\setboolean{useColorMapping}{true} 

\definecolor{bgYellow}{RGB}{255,245,170}
\definecolor{txtYellow}{RGB}{180,130,0}

\definecolor{bgRed}{RGB}{255,210,210}
\definecolor{txtRed}{RGB}{160,0,0}

\definecolor{bgGreen}{RGB}{210,255,210}
\definecolor{txtGreen}{RGB}{0,120,0}

\definecolor{bgBlue}{RGB}{210,230,255}
\definecolor{txtBlue}{RGB}{0,60,160}

\newcommand{\MaybeHL}[3]{%
  \ifthenelse{\boolean{useColorMapping}}%
    {{\sethlcolor{#1}\textcolor{#2}{\hl{#3}}}}%
    {#3}%
}

\newcommand{\YellowHL}[1]{\MaybeHL{bgYellow}{txtYellow}{#1}}
\newcommand{\RedHL}[1]{\MaybeHL{bgRed}{txtRed}{#1}}
\newcommand{\GreenHL}[1]{\MaybeHL{bgGreen}{txtGreen}{#1}}
\newcommand{\BlueHL}[1]{\MaybeHL{bgBlue}{txtBlue}{#1}}

\copyrightyear{2026}
\acmYear{2026}
\setcopyright{cc}

\acmConference[CHI EA '26]{Extended Abstracts of the 2026 CHI Conference on Human Factors in Computing Systems}{April 13--17, 2026}{Barcelona, Spain}
\acmBooktitle{Extended Abstracts of the 2026 CHI Conference on Human Factors in Computing Systems (CHI EA '26), April 13--17, 2026, Barcelona, Spain}
\acmDOI{10.1145/3772363.3798862}
\acmISBN{979-8-4007-2281-3/2026/04}

\begin{document}

\title[Bridging Pedagogy and Play]{Bridging Pedagogy and Play: Introducing a Language Mapping Interface for Human-AI Co-Creation in Educational Game Design}

\author{Daijin Yang}
\affiliation{%
  \institution{Northeastern University}
  \streetaddress{360 Huntington Ave.}
  \city{Boston}
  \state{Massachusetts}
  \country{USA}
  \postcode{02115}
}
\email{yang.dai@northeastern.edu}
\author{Erica Kleinman}
\affiliation{%
  \institution{Northeastern University}
  \streetaddress{360 Huntington Ave.}
  \city{Boston}
  \state{Massachusetts}
  \country{USA}
  \postcode{02115}
}
\email{e.kleinman@northeastern.edu}
\author{Casper Harteveld}
\affiliation{%
  \institution{Northeastern University}
  \streetaddress{360 Huntington Ave.}
  \city{Boston}
  \state{Massachusetts}
  \country{USA}
  \postcode{02115}
}
\email{c.harteveld@northeastern.edu}

\renewcommand{\shortauthors}{Trovato and Tobin, et al.}

\begin{abstract}
Educational games can foster critical thinking, problem-solving, and motivation, yet instructors often find it difficult to design games that reliably achieve specific learning outcomes. Existing authoring environments reduce the need for programming expertise, but they do not eliminate the underlying challenges of educational game design, and they can leave non-expert designers reliant on opaque suggestions from AI systems. We designed a controlled natural language framework-based web tool that positions language as the primary interface for LLM-assisted educational game design. 
In the tool, users and an LLM assistant collaboratively develop a structured language that maps pedagogy to gameplay through four linked components. We argue that, by making pedagogical intent explicit and editable in the interface, the tool has the potential to lower design barriers for non-expert designers, preserves human agency in critical decisions, and enables alignment and reflections between pedagogy and gameplay during and after co-creation.

\end{abstract}



\maketitle

\section{Introduction}

Educational games can enrich education by supporting critical thinking, problem solving, and collaboration in interactive learning environments~\cite{benefitsig1,enhancelearning1,ct1,problemsolving1,collaboration1}. Empirical studies further show that well-designed educational games can increase learners’ motivation and retention of disciplinary content~\cite{benefitsig1,benefitmotivation1}.

However, educational games often fall short when game experiences do not align with course content, assessment practices, or classroom routines, which can reduce pedagogical value and introduce conceptual confusion~\cite{barrier1,barrier2,contextbarrier}. These misalignments contribute to instructors’ continued hesitancy to adopt educational games systematically~\cite{differencek12andhigher1,lowadoption}. Although instructors understand learning objectives and student needs, many lack training in game and learning experience design, and they face persistent challenges in configuring mechanics, goals, narrative, and user experience to serve specific learning outcomes without sustained scaffolding~\cite{teacherknow1,barrier4,understanddesign1,barriersgamecreation1}. Low-code authoring environments can lower programming barriers, but they do not resolve these deeper design and game-literacy demands~\cite{emergo2,uadventure,studycrafter,understanddesign1,barriersgamecreation1}.

Large Language Models (LLM) offer a possible response by assisting instructors across design stages, from early ideation to prototype implementation, and by enabling mixed-initiative co-creative workflows~\cite{lowadoption,gpt4games2,mixedinitiativegamereview,mixedinitiativeinterface,cocreativity}. By generating candidate content, code, and prototypes from natural language or structured design documents, these tools can reduce manual programming and help instructors translate pedagogical intentions into playable experiences~\cite{gpt4games2}.

However, naive use of LLM assistants introduces risks for instructors. An instructor may receive a seemingly complete game with limited insight into how it supports stated learning objectives, which content elements are fabricated, or how design choices may exacerbate challenges for students who already struggle with gameplay or disciplinary understanding~\cite{contextbarrier,studentbarrier,barrier1}. Prior work on LLM-supported design cautions that opaque automation can diminish human agency by forcing instructors to reverse engineer under-specified suggestions or by relegating them to content approvers rather than designers of educational intent~\cite{mixedinitiativegamereview,cocreativity}. Consequently, AI-supported educational game design requires communication that foregrounds transparency and fine-grained control, enabling instructors to inspect, critique, and iteratively adjust AI proposals while maintaining alignment with course goals and diverse learner needs~\cite{barrier4,teacherknow1,teacherknow2}.

We address these needs by designing a web interface that operationalizes a language-based framework as the central collaboration interface for LLM-assisted educational game design. Instructors and the system co-author a short, structured description of a learning activity, and the interface generates a corresponding game-level description using the same sentence structure. By making this shared representation editable and inspectable, the tool links pedagogical intent to concrete game decisions about rules, content, difficulty, and context, while providing stable handles for constraint setting, explanation, and revision. In doing so, the tool shows the potential to instantiate a controlled language-mediated communication protocol for human--AI co-creation that supports goal negotiation, traceability, and responsibility in generative design workflows.

\section{Related Work}
Prior work establishes both the promise and the practical difficulty of educational games, and it motivates approaches that embed Large Language Models (LLM) within transparent, instructor-centered co-creative processes. Although educational games can support motivation and learning through interactive practice and feedback~\cite{benefitsig1,subhash2018review,benefitmotivation1}, instructors often hesitate to adopt them when game experiences misalign with course content, assessment practices, and classroom routines, or when suitable resources and institutional support are limited~\cite{contextbarrier,barrier1,barrier2,barrier3,lowadoption}. Even when instructors wish to create or adapt games, they frequently lack preparation in game and learning experience design, and they must fit design work into constrained instructional time~\cite{teacherknow1,teacherknow2,barrier4}. 

Recent LLM-based game design tools appear to lower technical barriers by generating ideas, rules, assets, and prototypes from prompts or structured descriptions~\cite{gatti2023chatgpt,swacha2025genaiserious,hu2024gamegen,todd2023level,gallotta2024llmgames}. However, studies of LLM-supported generation also highlight persistent concerns about limited controllability and transparency, which can shift effort toward verification and repair to address incoherence, inaccuracies, or weak pedagogical alignment~\cite{todd2023level,hu2024gamegen,swacha2025genaiserious,spasic2025chatgptgame}. Educators similarly report concerns about hallucinations, misalignment with objectives, and increased checking workload, which can undermine pedagogical control even when productivity improves~\cite{hasanein2023drivers,park2024promise,ogurlu2023perception}. Human--AI co-creation frameworks help structure collaboration through roles, phases, and initiative, but they often provide limited guidance on how instructors should communicate intent, constraints, and rationales to support inspection and revision during design~\cite{chaitMcGrath2024,ciloopSmith2022,ucccFrameworkJones2023,cofiRezwanaMaher2023,faicoRezwanaFord2025}. 

As one of the possible solutions, controlled natural languages demonstrate how constrained, readable text can reduce ambiguity while remaining machine parsable, enabling domain experts to author, inspect, and revise specifications with computational support~\cite{kuhn2014cnl,fuchs2008attempto}. Analogous language-oriented representations in game design, including the Video Game Description Language and the Learning Mechanics--Game Mechanics model, further show a potential way how structured descriptions can expose design choices and connect mechanics to learning processes in ways that support reflection and iteration~\cite{vgdlSchaul2013,lmgmArnab2015,proulx2018lmgmsdt}. Taken together, the literature motivates instructor-centered co-creative tools that use language-mediated representations to make pedagogical intent explicit, inspectable, and revisable while leveraging LLM generation.

\section{Introducing the educational game design language mapping framework}

Based on the literature, we propose a language mapping based framework. The framework represents educational intent as a single controlled sentence that is readable to humans and parseable by the system, and acts as the co-creation interface:

\begin{quote}
``Players (Students) \textbf{[\textit{\textbf{\RedHL{Adverbs}}}] [\textit{\textbf{\YellowHL{Verbs}}}] [\textit{\textbf{\GreenHL{Nouns}}}] in a [\textit{\textbf{\BlueHL{Adjectives}}}]} environment.''
\end{quote}

This template is the basic unit of specification. It partitions intent into four grammatical elements, each carrying a distinct semantic role, and fixes their positions so that the same surface form can serve as a shared reference for both pedagogical description and system interpretation. Table~\ref{tab:edu-game-language-mapping} summarizes how each element maps between teaching language and game language, establishing an explicit correspondence between instructional meaning and game-design implications.

Within the framework, \textit{\textbf{\YellowHL{Verbs}}} define the targeted ability as an observable learning action, and they serve as the core semantic anchor for selecting a family of player interactions. \textit{\textbf{\GreenHL{Nouns}}} define the focal concept or content domain, constraining what the activity is about and what must be represented. \textit{\textbf{\RedHL{Adverbs}}} define performance requirements, expressing what counts as adequate enactment of the targeted ability and thereby delimiting difficulty and success conditions. \textit{\textbf{\BlueHL{Adjectives}}} define contextual and stylistic constraints by characterizing the learning environment, including the realism level and instructional tone, which, in turn, determine the appropriate aesthetic and narrative framing on the game side.

Taken together, the four elements form a compact specification that keeps pedagogical intent explicit while making its design consequences legible. Each element corresponds to a manipulable handle in the representation: abilities (\textit{\textbf{\YellowHL{Verbs}}}), content (\textit{\textbf{\GreenHL{Nouns}}}), performance criteria (\textit{\textbf{\RedHL{Adverbs}}}), and context or tone (\textit{\textbf{\BlueHL{Adjectives}}}). By requiring all four to be present in each sentence, the framework encourages complete descriptions that can be interpreted consistently and compared across iterations.

\begin{table*}[t]
\centering
\caption{Mapping between teaching language and game language in the educational game design framework.}
\label{tab:edu-game-language-mapping}
\renewcommand{\arraystretch}{1.2}
\small
\begin{tabularx}{\textwidth}{
  >{\raggedright\arraybackslash}p{0.12\textwidth}
  >{\raggedright\arraybackslash}p{0.38\textwidth}
  >{\centering\arraybackslash}p{0.06\textwidth}
  >{\raggedright\arraybackslash}X
}
\toprule
\multicolumn{1}{>{\centering\arraybackslash}p{0.12\textwidth}}{%
  \parbox[c][5ex][c]{0.12\textwidth}{\centering\textbf{Grammatical elements}}%
} &
\multicolumn{1}{>{\centering\arraybackslash}p{0.38\textwidth}}{%
  \parbox[c][5ex][c]{0.38\textwidth}{\centering\textbf{Teaching language}}%
} &  &
\multicolumn{1}{>{\centering\arraybackslash}X}{%
  \parbox[c][5ex][c]{\hsize}{\centering\textbf{Game language}}%
} \\
\midrule
\textit{\textbf{\RedHL{Adverb}}} &
Specifies performance requirements for the targeted ability. &
$\longrightarrow$ &
Rules and parameters that configure difficulty and success conditions. \\
\midrule
\textit{\textbf{\YellowHL{Verbs}}} &
Expresses the targeted teaching ability as an observable action. &
$\longrightarrow$ &
Game mechanics that define the primary player action and interaction pattern. \\
\midrule
\textit{\textbf{\GreenHL{Nouns}}} &
Denotes the focal teaching concept or content domain. &
$\longrightarrow$ &
Content models and in-game artifacts that instantiate the concept. \\
\midrule
\textit{\textbf{\BlueHL{Adjectives}}} &
Characterizes the learning context, realism level, and instructional tone. &
$\longrightarrow$ &
Aesthetic and contextual profiles that define the game world and framing. \\
\bottomrule
\end{tabularx}
\end{table*}

\section{System Design}
Based on the framework, we further designed and developed an LLM-based web application to support instructors in designing educational games. In the application, the instructors go through 3 phases: the requirement extraction phase, where the AI asks several questions to extract the key grammatical elements defined in the last section; the translation phase, where the AI helps to translate the abstracted teaching language sentence into game language sentence; and the language development phase, where the AI helps to expand the translated sentence into more detailed descriptive paragraphs.

\subsection{Requirement Extraction}
\begin{figure*}[!htbp]
\centering 
\includegraphics[width=0.85\textwidth]{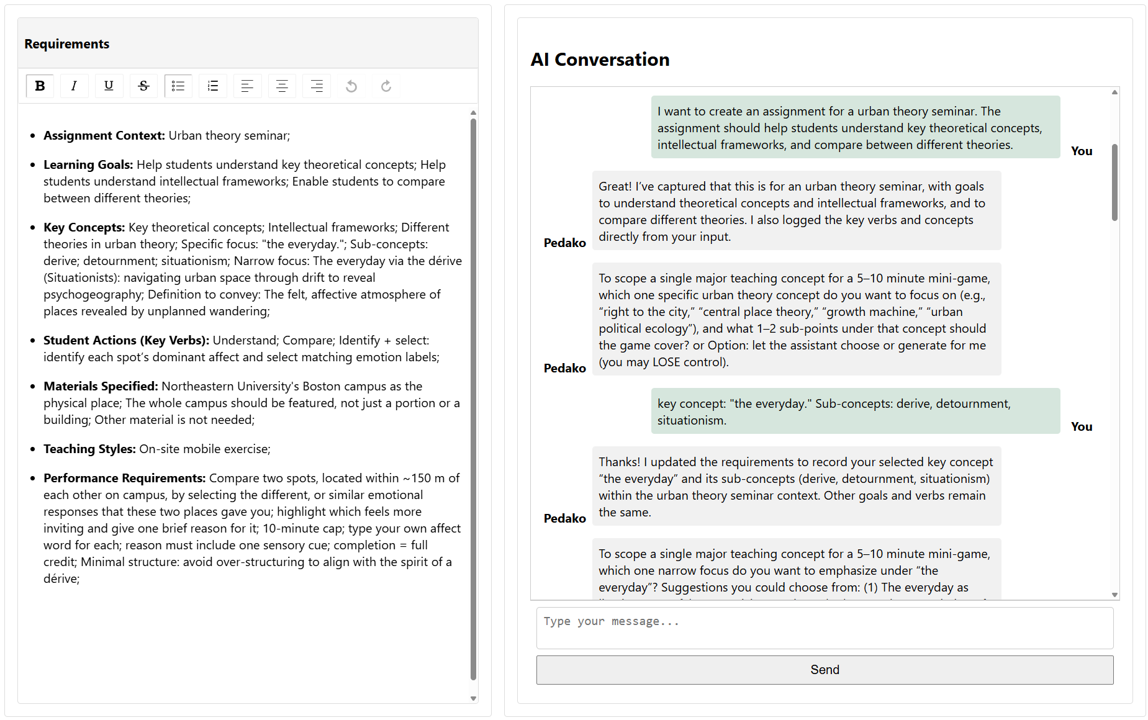} 
\caption{The Requirement Extraction Components in the Tool. Instructors answers questions asked by AI around the defined grammatical elements on the right and co-author the requirement document on the left.}
\label{fig:req}
\end{figure*}

As shown in Fig.~\ref{fig:req}, after the instructor states an initial goal, the AI assistant runs a structured dialogue to elicit the four pedagogical grammatical elements defined by the framework: the focal concept and materials (\textit{\textbf{\GreenHL{Nouns}}}), the observable target ability (\textit{\textbf{\YellowHL{Verbs}}}), the performance requirement that sets difficulty and success conditions (\textit{\textbf{\RedHL{Adverbs}}}), and the instructional context that defines situational framing (\textit{\textbf{\BlueHL{Adjectives}}}).

For each component, the assistant asks guidance questions, checks whether responses are specific enough to translate into design constraints, and proposes a small set of candidate options to support refinement through selection and revision. It asks the instructor to specify (1) the intended concept or skill at an appropriate scope, (2) the content resources or materials that should be used, (3) the learner action that should be observable by the end of the activity, (4) measurable performance targets that define success within the intended time window, and (5) contextual requirements such as environment type, realism level, and instructional tone. In this phase, instructors are granted maximum authority to shape the design in ways they are familiar with. They can also reflect on their teaching through a game-language lens, using a shared language structure from game language to shape the teaching language.

\subsection{Language Translation}
\begin{figure*}[!htbp]
\centering 
\includegraphics[width=0.85\textwidth]{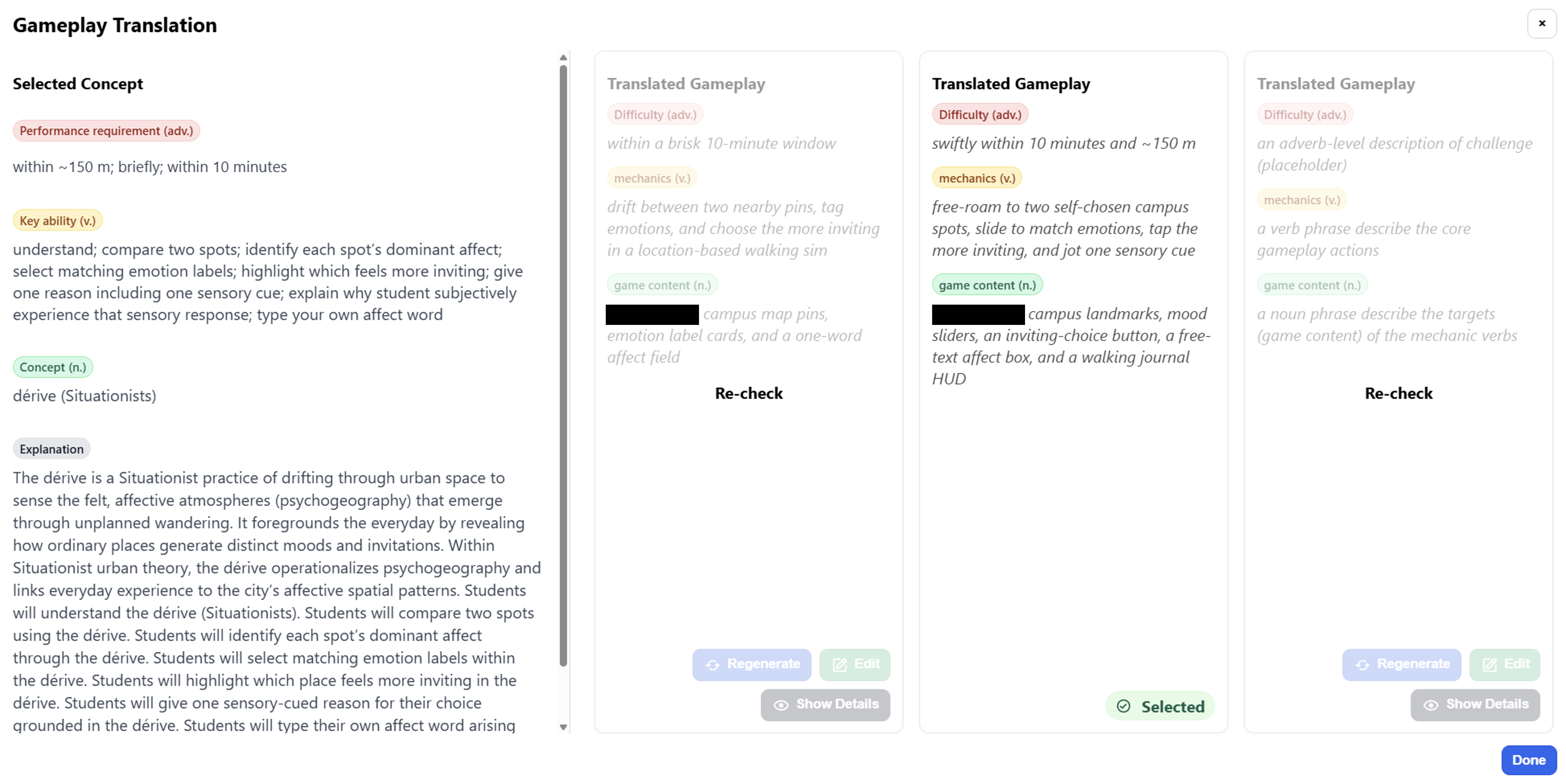} 
\caption{The Language Translation Components (Partially) in the Tool. Instructors explore AI-generated gameplay translations of their teaching language. They can edit, regenerate, or write their own translations during the process. The interface uses color mapping to link corresponding grammatical elements in the teaching language and the game language.}
\label{fig:trans}
\end{figure*}

The instructor’s answers constrain the assistant to compose a single pedagogy sentence that fills the four slots of the template (\textit{\textbf{\RedHL{Adverbs}}}, \textit{\textbf{\YellowHL{Verbs}}}, \textit{\textbf{\GreenHL{Nouns}}}, \textit{\textbf{\BlueHL{Adjectives}}}). 
As shown in the left part of Fig.~\ref{fig:trans}, the slots function as a compact specification of the intended learning activity, capturing the targeted ability, focal content, performance requirements, and contextual framing without requiring additional formal notation. 

Given this structured pedagogy language, as shown in the right part of Fig.~\ref{fig:trans}, the assistant then generates multiple candidate game-language translations that satisfy the same constraints. Each candidate expresses a game as a parallel four-part specification aligned with the framework: difficulty and success conditions (\textit{\textbf{\RedHL{Adverbs}}}), mechanics that operationalize the targeted ability (\textit{\textbf{\YellowHL{Verbs}}}), game content that instantiates the focal concept and materials (\textit{\textbf{\GreenHL{Nouns}}}), and aesthetic or setting descriptors that instantiate the instructional context (\textit{\textbf{\BlueHL{Adjectives}}}). For each candidate, the assistant provides a component-wise rationale that explains how each game-side element corresponds to the pedagogy-side specification. This makes alternatives comparable at the level of individual slots, supporting targeted revision when a proposal does not satisfy constraints such as time-on-task, allowable support, representational format, or contextual framing. 

For each candidate, the assistant provides a component-wise rationale that explains how each game-side element corresponds to the pedagogy-side specification. By enabling the instructor to compare, revise, and commit to alternative translations, this phase supports tracing each part of the design decision-making process and maintains transparency through an inspectable, traceable mapping between pedagogy language and gameplay language that the instructor can refine.

\subsection{Language Development}
\begin{figure*}[!htbp]
\centering 
\includegraphics[width=0.85\textwidth]{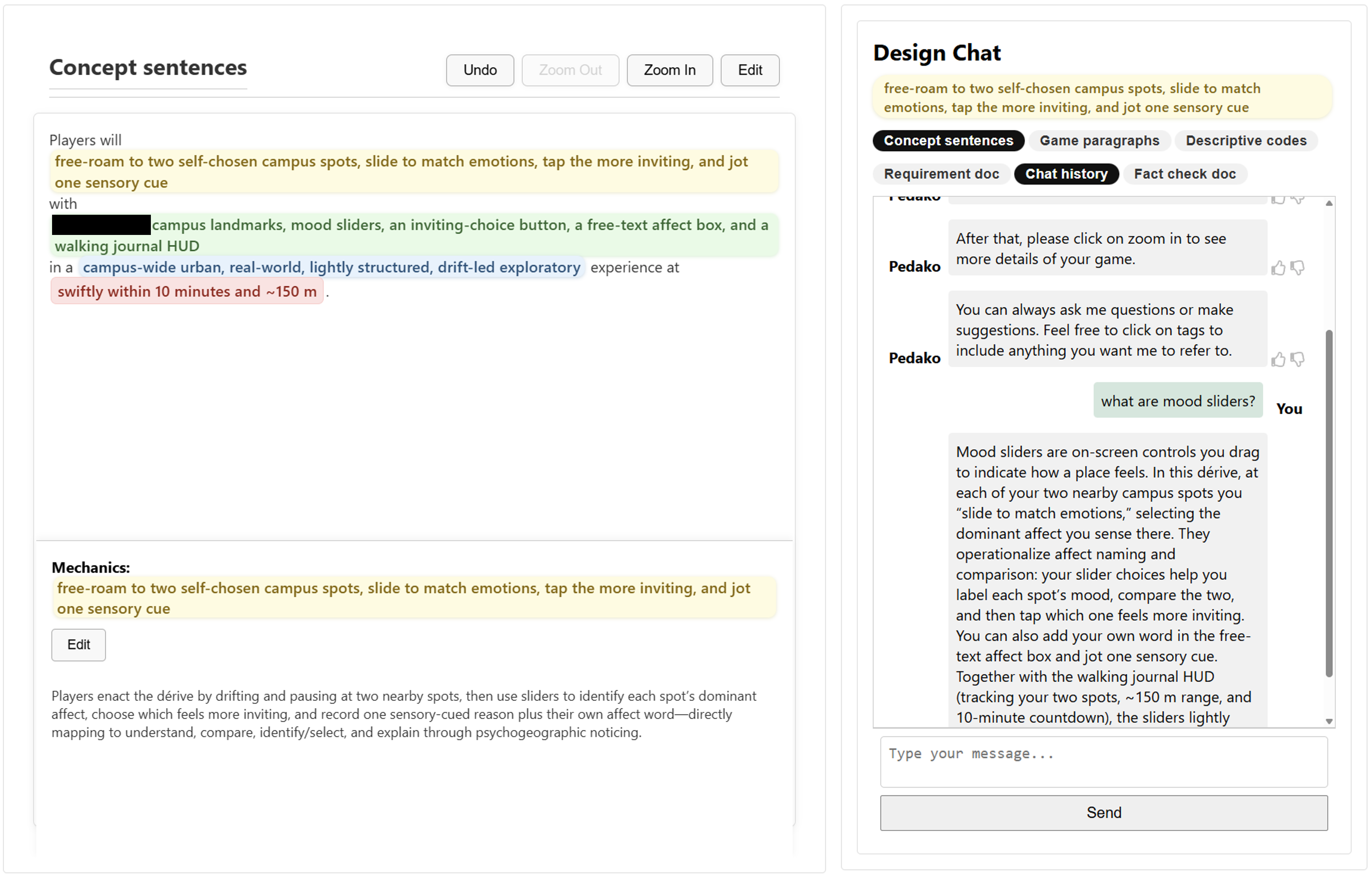} 
\caption{The Language Development Components (Partially) in the Tool. Instructors can further edit and refine the generated game-concept sentence with AI. They can also click the Zoom In buttons to have the AI generate a descriptive paragraph and pseudocode for the game’s procedures.}
\label{fig:dev}
\end{figure*}

After instructors complete the translation phase, as shown in Fig.~\ref{fig:dev}, they can further edit and refine the game-language sentence with support from the AI assistant. Their chats with the AI assistant are context-aware. They can ask the AI to clarify, change, or regenerate any part of the sentence. They can also click the Zoom In buttons to ask the AI to generate descriptive paragraphs with play examples that help instructors better understand how the game operates. If they click the Zoom In buttons again, the tool generates pseudocode for the game, which instructors can use as a reference when developing the game with other tools. Because our tool focuses on supporting educational game design phases, the final output is the pseudocode instead of a real game.

\subsection{Implementations}
We implemented the proposed system as a full-stack web application that supports the three-phase workflow described above and preserves instructor inputs, intermediate representations, and generated outputs for iterative revision. The application stores project data, including requirement responses, pedagogy-language and game-language sentences, and versioned artifacts, in a MongoDB database. The front end is built with TypeScript and Node.js to provide an interactive interface with color-coded slot alignment and in-place editing across phases. The back end is implemented in Python and handles data orchestration, prompt construction, logging, and calls to the AI service. For AI generation, the system uses the GPT-5 Application Programming Interface (API)~\cite{GPTDocument} without fine-tuning, relying instead on structured prompting that constrains outputs to the controlled template and phase-specific objectives.
\section{Conclusions and Future Studies}

We presented a language-based co-creation interface for LLM-assisted educational game design that uses a controlled sentence template to make pedagogical intent explicit, editable, and traceable throughout co-creation. The web tool structures instructor–AI collaboration across requirement extraction, language translation, and language development, enabling instructors to iteratively refine both teaching-language and game-language representations while preserving fine-grained control. The interface supports comparison, revision, and reflection by maintaining an inspectable mapping between pedagogy language and gameplay language, culminating in pseudocode as an actionable design output for downstream development. 

For future work, we plan to evaluate the tool with instructors in authentic course-planning contexts, test how the representation supports pedagogical alignment and reflection across iterative design cycles, and study how pseudocode outputs integrate with existing educational game authoring workflows. We will also examine the interaction dynamics of instructor–LLM co-creation across phases, including how instructors initiate, negotiate, accept, and revise suggestions, and how these patterns relate to trust calibration and mental model formation over time. In addition, we will assess instructors’ post-interaction personal gains by characterizing perceived ownership, self-efficacy, and knowledge gains after the co-created design process. Finally, we will conduct an A/B evaluation that compares the proposed language-mapping framework against a baseline condition that uses standard LLM prompting for the same design tasks. Collectively, these studies will clarify when and why the framework improves design outcomes, and how it shapes interaction trajectories in LLM-supported educational game design.

\begin{acks}
This work was supported by the National Science Foundation (NSF) under Grant \#2142396. Any opinions, findings, and conclusions or recommendations expressed in this material are those of the author(s) and do not necessarily reflect the views of the National Science Foundation.
\end{acks}

\bibliographystyle{ACM-Reference-Format}
\bibliography{references}

\appendix

\end{document}